\documentclass[aps,pra,nofootinbib,11pt,notitlepage]{revtex4-1}
\usepackage{graphicx}
\usepackage{multirow}
\usepackage{amsmath}
\usepackage{amssymb}
\usepackage{amsthm}
\usepackage{eucal}
\usepackage{bm}
\usepackage{upgreek}
\usepackage{framed}
\usepackage{mathrsfs}
\usepackage{latexsym}
\usepackage{float}
\usepackage{subfigure}
\usepackage{color}
\usepackage{bm,hyperref,tikz}
\usepackage[utf8]{inputenc} % Use utf-8 encoding for foreign characters
\usepackage{bbold}
\usepackage{bbm}
\usetikzlibrary{decorations.pathmorphing, patterns,shapes}

\newcommand{\crb}{Cram\'{e}r-Rao bound}
\newcommand{\ada}{a^{\dagger}a}
\newcommand{\bdb}{b^{\dagger}b}

\begin{document}
	\title{Two-mode Gaussian product states in a Lossy Interferometer }
	\author{Noufal Jaseem}
	\email{noufaljaseemp@iisertvm.ac.in}
	\author{Anil Shaji}
	\email{shaji@iisertvm.ac.in}
	\affiliation{School of Physics, Indian Institute of Science Education and Research Thiruvananthapuram, Kerala  695016, India}
	
	\begin{abstract}
		The quantum Fisher information for a two-mode, Gaussian product state in an interferometer subject to photon loss is studied. We obtain the quantum Cramer-Rao bound on the achievable precision in phase estimation using such states. The scaling of the measurement precision with the mean photon number for such input product states is compared to the  limited scaling for dual squeezed vacuum states and for dual squeezed, displaced vacuum states. 
	\end{abstract}
	\maketitle
	%%%%%%%%%%%%%%%%%%%%%%%%%%%%%%%%%%%%%%%%%%%%%%%%%%%%%%%%%%%%%%%%%%%%%%%%%%%%%%
	\section{Introduction}
	
	Using light as the measuring device by employing interferometric arrangements have led to a long list of fundamentally significant advances in our understanding of the universe ranging from the negative result of the Michaelson-Morely experiment~\cite{Michelson:1887ka} to the recent detection of gravitational waves~\cite{abbott2016observation,Adhikari:2014gx}. There is ample evidence to suggest that by tapping into the nonClassical features of light it is possible to push the limits of precision interferometry further~\cite{Caves:1980vh,demkowicz2015chapter,Taylor20161review,DemkowiczDobrzanski:2009gl,dorner2009optimal,dowling08a,Fupper,lang2014optimal,sahota2015quantum,vsafranek2015quantum,zhang2013lossy,Adesso:2014hx,Anisimov:2010kz,Aspachs:2009fu}. The varieties of quantum features of states of light and the possibilities presented by them are virtually limitless. Experimental investigations of many of them are also becoming possible in recent times.  
	
	The lowest attainable bound on the measurement uncertainty given the resources available to perform the measurement is the central question in metrology. Quantum features of light in an interferometer is the resource of interest in this Paper. The measured parameter in a typical interferometer is a phase difference $\varphi$ and the scaling of the uncertainty in the estimated phase, $\delta \varphi$ with respect to the mean number of photons, $\bar{N}$, in the state of light that is the input to the interferometer is the quantity that is focused on here. With `classical' states of light like coherent states as input to an interferometer  $\delta \varphi$ scales with the mean number of photons as $\delta \varphi\geq 1/\sqrt{\bar{N}}$ and this is the standard quantum limit (SQL) or the shot noise limit. If one were to devise highly non-classical states of light, like N00N and squeezed vacuum states, one can do better, in principle, than the SQL~\cite{dowling08a,lang2014optimal}. With N00N or Squeezed Vacuum states  Heisenberg-limited scaling of $\delta \varphi\geq 1/\bar{N}$ can be achieved. However, many of these well-studied non-classical states yield Heisenberg-limited precision only in the absence of decoherence, photon loss, and other noise. In any practical interferometer, noise is inevitable. N00N states are highly sensitive and extremely fragile to the photon losses and are outperformed by classical states of light in the presence of losses~\cite{Knysh:2011ed,Koiodynski:2010cg,dorner2009optimal}. 
	
	The performance of various quantum states of light with fixed photon number in interferometers, as well as the effect of photon loss and other forms of decoherence in such states, have been studied previously ~\cite{dorner2009optimal,DemkowiczDobrzanski:2009gl,Koiodynski:2010cg,Knysh:2011ed,Escher:2011fn,DemkowiczDobrzanski:2012gl,PhysRevA.92.032112}. As discussed in the following, fixing the mean photon number as the measure of the resources does not unambiguously fix the best possible scaling of the measurement precision, and so an additional constraint is required for states other than those with fixed photon numbers. In~\cite{lang2014optimal} the additional logical constraint that the input state is a product state across the two input ports of a typical Mach-Zehnder interferometer is introduced. In the absence of losses, the dual squeezed vacuum state (DSV) is shown to be optimal input product state leading to the best possible scaling of $\delta \varphi$ with respect to $\bar{N}$. In this paper we consider optimal input two-mode Gaussian product states into a Mach-Zehnder type interferometer in the presence of photon loss
	
	In section~\ref{sec:fisher} we review the quantum Fisher information for  two-mode Gaussian states. The quantum Fisher information for arbitrary two-mode Gaussian product states is obtained in section~\ref{sec:gaussian}. We see that in the presence of photon losses these states as input to interferometers do not lead to better scaling for the measurement uncertainty with respect to the mean photon number compared to other `classical' states of light. The enhancement in the measurement precision is only through constant factors. In section~\ref{sec:special}, we study the DSV and dual squeezed, displaced vacuum (DSDV) states as special cases. Section V summarises our findings. 
	
	%%%%%%%%%%%%%%%%%%%%%%%%%%%%%%%%%%%%%%%%%%%%%%%%%%%%%%%%%%%%%%%%%%%%%%%%%%%%%%%%%
	
	\section{The quantum Fisher Information \label{sec:fisher}}
	
	The quantum state of light in an interferometer is the particular part of the measuring device that responds directly to small changes in the measured quantity. The measurement uncertainty is inversely proportional to the magnitude of the response of the measuring device to small changes in the measured quantity. This statement in made precise, when the relevant part of the measuring device is a quantum state, $\rho$, by the quantum \crb~\cite{braunstein94a, braunstein96a, helstrom76a, holevo82a} as
	\begin{equation}
		\label{eq:crb}
		\Delta \varphi \geq \frac{1}{\sqrt{{\cal I}(\varphi)}},
	\end{equation}
	where ${\mathcal I}(\varphi)$ is the {\em quantum Fisher information}. 
	
	The $\varphi$ dependent part of the dynamics of the state in the interferometer usually consists of two parts, unitary evolution, and non-unitary evolution.  We assume that the unitary part of the time evolution of the state of light in the interferometer (the signal) is generated by a Hamiltonian of the form 
	\begin{equation}
		\label{eq:probeH}
		H_{\varphi} = \varphi H,
	\end{equation}
	Non-unitary evolution is relevant for any practical interferometric with noise. In the basis $\{|j\rangle \}$ in which $\rho$ is diagonal ($\rho = \sum_{j} p_{j} |j\rangle \langle j|$), the Fisher information has the form \cite{braunstein94a}
	\begin{eqnarray}
		\label{eq:I0}
		{\cal I}(\varphi)= \sum_j \frac{\big(dp_j/d\varphi\big)^2}{p_j}+2\sum_{j,k} \frac{(p_j-p_k)^2}{p_j+p_k} |H_{jk}|^2, \qquad H_{jk} = \langle j | H | k \rangle. 
	\end{eqnarray}
	In the above equation, the first term comes from the non-unitary, noise part of the evolution and the second term comes from the unitary part. The expression in Eq.~(\ref{eq:I0}) for the quantum Fisher information is not always easy to handle, especially when the state is not readily diagonalisable. The two-mode Gaussian states we study are not easy to diagonalise, especially in the presence of decoherence. A general expression for the quantum Fisher information of two-mode Gaussian states is derived in~\cite{vsafranek2015quantum}, which we outline briefly below.
	
	The quantum Fisher information is also a measure of how well we can distinguish two neighbouring states $\rho_{\varphi}$ and $\rho_{\varphi+\epsilon}$ and it can be defined as the limit~\cite{hayashi2006quantum},
	\begin{equation}\label{eq:I1}
		\mathcal{I}(\varphi)=8 \lim_{\epsilon \rightarrow 0} \frac{1-\sqrt{\mathcal{F}(\rho_\varphi,\rho_{\varphi+\epsilon})}}{\epsilon^2},
	\end{equation}
	where $\mathcal{F}(\rho_\varphi,\rho_{\varphi+\epsilon})$ is the fidelity defined as  		
	\begin{equation}
		\label{fid}
		\mathcal{F}(\rho_\varphi,\rho_{\varphi+\epsilon}) = \Big[{\rm Tr} \Big(\sqrt{\sqrt{\rho_\varphi}			\rho_{\varphi+\epsilon} \sqrt{\rho_\varphi}} \,\Big)\Big]^2.
	\end{equation}
	The Fidelity of a two-mode Gaussian state~\cite{marian2012uhlmann} is given by
	\begin{equation}
		\mathcal{F}(\rho_{\varphi},\rho_{\varphi+\epsilon})= 
		\frac{4 e^{\delta \vec{d}^\dagger (\Sigma_{\varphi}+\Sigma_{\varphi+\epsilon})^{-1} \delta \vec{d}}}{\sqrt{\Gamma}+\sqrt{\Lambda}-\sqrt{(\sqrt{\Gamma}+\sqrt{\Lambda})^2-\Delta}}.
	\end{equation}
	The Fidelity is given in terms of the expectation values and covariances of the creation and annihilation operators of the two modes ($a$, $a^{\dagger}$) and ($b$, $b^{\dagger}$) respectively. In the expression above, $\vec{d} = \langle \vec{u} \rangle$, where $\vec{u}=(a,b,a^\dagger,b^\dagger)^T$ and $\delta \vec{d} = \vec{d}_{\varphi} - \vec{d}_{\varphi + \epsilon}$. $\Sigma$ is the covariance matrix with elements 
	\begin{equation}
		\label{eq:covar}
		\Sigma_{ij}=\langle \{(u_i-\langle u_i\rangle),(u_j-\langle u_j\rangle)^\dagger\}\rangle, 
	\end{equation}
	where  $\{.,.\}$ and $\langle . \rangle$ denote the anti-commutator and the expectation value respectively. Further, we have
	\begin{eqnarray}
		\Delta &= & |\Sigma_\varphi+\Sigma_{\varphi+\epsilon}|, \nonumber \\
		\Gamma &= &|\mathbbm{1}+K\Sigma_{\varphi}K\Sigma_{\varphi+\epsilon}|, \nonumber \\
		\Lambda &= & |\Sigma_{\varphi}+K| |\Sigma_{\varphi+\epsilon}+K|,  
	\end{eqnarray}
	where 
	\[ K =  \openone \oplus -\openone = \left[ \begin{array}{cc} \openone & 0 \\ 0 & -\openone \end{array} \right]. \] 
	
	As detailed in~\cite{vsafranek2015quantum} we can Taylor expand the fidelity upto second order in $\epsilon$  around the point $\varphi$ to compute the limit in Eq.~(\ref{eq:I1}) and the Quantum Fisher Information can be written as 
	\begin{eqnarray}
		\label{eq:I2}
		\mathcal{I}&=& \frac{1}{2(|A|-1)}  \bigg\{ |A| {\rm tr} \Big[ \big(A^{-1}\dot{A}\big)^2\Big]+ \sqrt{|\mathbbm{1}+A^2|}  {\rm tr} \Big[\big( ( \openone +A^2)^{-1}\dot{A}\big)^2\Big] \nonumber\\
		& & \quad +\; 4\big(\tau_1^2 -\tau_2^2 \big) \bigg( \frac{\dot{\tau}_2^2}{\tau_2^4-1}-\frac{\dot{\tau}_1^2}{\tau_1^2-1} \bigg) \bigg\} + 2 \dot{\vec{d}}^\dagger \Sigma^{-1} \dot{\vec{d}},
	\end{eqnarray}
	where $A=K\Sigma$, $\tau_i$'s are symplectic eigenvalues of $\Sigma$ and the dots above the symbols denote the derivative with respect to $\varphi$. The symplectic eigenvalues of $\Sigma$ are obtained as,
	\[ \tau_{1,2} = \frac{1}{2} \sqrt{{\rm tr} \big( A^{2} \big) \pm \sqrt{ \big[{\rm tr} \big(A^{2} \big) \big]^{2} - 16|A|}}, \]
	where $|\cdot |$ denotes the determinant. Eq.~(\ref{eq:I2}) is applicable to all two-mode Gaussian states. The states we study in the Paper are Gaussian states and the noise that they are subjected to preserve the Gaussian nature and so this expression for the quantum Fisher information is applicable. 
	
	\subsection{Mean photon number and the quantum Fisher information}
	
	At this point, we digress a bit and note that the generator of the relative phase between the arms of interferometers is usually proportional to the photon number operator, $H \propto \ada = \hat{N}$. This means that the minimum measurement uncertainty in phase estimation using a pure state in a lossless interferometer is inversely proportional to the variance in $\hat{N}$ as $\delta \varphi \geq \frac{1}{2 \sqrt{\Delta H}}$ for pure states. This presents a problem when we seek to use the mean photon number $\bar{N}$ to quantify the resources put into a measurement scheme because it is possible to keep the mean fixed and increase the variance of $N$ arbitrarily. 
	
\begin{figure}[!htb]
		\begin{tikzpicture}
		%%%%%%%%%%%%%%%%%%%%%% Upper line
		\draw[gray!80!white,very thick] (-5.5,1.8) -- (-4.3,1.8) -- (-3,-1.8) -- (-1.6,-1.8);
		\draw[gray!80!white,very thick] (-0.6,-1.8) -- (0,-1.8);	
		%%%%%%%%%%%%%%%%%%%% Lower Line	
		\draw[gray!80!white,very thick] (-5.5,-1.8) -- (-4.3,-1.8) -- (-3,1.8)-- (-1.6,1.8);
		\draw[gray!80!white,very thick] (-0.6,1.8) -- (0,1.8);	
		%%%%%%%%%%%%%%%%%%%%%%%% Beam splitter
		\draw[line width=4pt, color=blue!30!white] (-4.0,0) -- (-3.3,0) node [anchor= south west, color=black] {$B$};
		%%%%%% upper beam splitter
		\draw[line width=4pt, color=blue!30!white] (-2.6,1.5) -- (-2.1,2.1) node [anchor= south west, color=black] {$\eta_{a}$};
		%\draw[gray!80!white,very thick] (-2.35,2.5) -- (-2.35,1.8);
		\draw[decoration = {zigzag,segment length = 4mm, amplitude = 0.5mm},decorate,gray!80!white,very thin] (-2.37,2.5) -- (-2.37,1.8);
		\draw[decoration = {zigzag,segment length = 4mm, amplitude = 0.5mm},decorate,gray!80!white,very thin] (-2.32,2.5) -- (-2.32,1.8);
		%%%%% Lower beam splitter	
		\draw[line width=4pt, color=blue!30!white] (-2.6,-2.1) -- (-2.1,-1.5) node [anchor= south west, color=black] {$\eta_{b}$};
		%\draw[gray!80!white,very thick] (-2.35,-1.9) -- (-2.35,-1.0);
		\draw[decoration = {zigzag,segment length = 4mm, amplitude = 0.5mm},decorate,gray!80!white,very thin] (-2.37,-1.8) -- (-2.37,-1.0);
		\draw[decoration = {zigzag,segment length = 4mm, amplitude = 0.5mm},decorate,gray!80!white,very thin] (-2.32,-1.8) -- (-2.32,-1);
		%%%%%%%%%%%%%%%%% labels \psi and a,b
		\draw[] (-6,1.8) node{$|\psi_{1}\rangle$};
		\draw[] (-6,-1.8) node{$|\psi_{2}\rangle$};
		\draw[] (-5,2) node{$a$};
		\draw[] (-5,-2.1) node{$b$};
		%%%%%%%%%%%%%%%%%%% label \varphi/2    
		\draw[] (-1.1,1.8) node{$\varphi/2$}circle (0.5 cm);
		\draw[] (-1.1,-1.8) node{$-\varphi/2$}circle (0.5 cm); 
		\fill[color=green!20] (0,1.9) rectangle (1, -1.9);
		\draw (0,1.9) rectangle (1, -1.9);
		\node[rotate=90] at (0.5, 0) {\large Measurement};
		\end{tikzpicture}
		\caption{An interferometric setup without a second beam splitter in symmetric configuration with phase shifts $+\varphi/2$ and $-\varphi/2$ in each of the two arms respectively. $B$ denotes beam splitter. There are two input ports labelled $a$ and $b$ and Gaussian states $|\psi_{1}\rangle$ and $|\psi_{2}\rangle$ are fed into modes $a$ and $b$ respectively. Losses are modelled by introducing additional beam splitters along each of the modes which removes a fraction of the photons from each mode as shown. \label{fig1}}
	\end{figure}
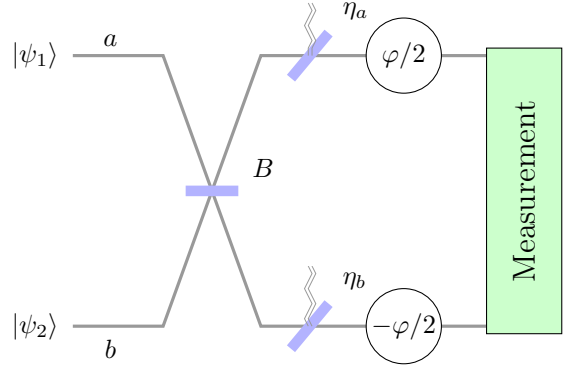

	For an interferometer in the symmetric configuration as shown in Fig.~\ref{fig1} but without losses, we have 
	\begin{equation}
		\label{eq:hamil2}
		H_{\rm interf} = \varphi H = \varphi \frac{1}{2}\big(a^{\dagger}a - b^{\dagger}b \big).
	\end{equation}
	
	Let the input state of light in the interferometer after the first beam splitter be an arbitrary, two mode, pure state,
	\begin{equation}
		\label{eq:genstate}
		|\Psi\rangle = \sum_{n_{1},n_{2}=0}^{\infty} c_{n_{1} n_{2}}|n_{1},n_{2}\rangle =\sum_{N=0}^{\infty}  \sum_{n=-N}^{N}C_{N,n}|N,n\rangle ,
	\end{equation}
	where $n_{1}$ and $n_{2}$ are the occupancies in each of the two modes, $N=n_{1}+n_{2}$ and $n=n_{1}-n_{2}$. Let
	\begin{equation}
		\label{eq:normgen}
		p_{N} \equiv \sum_{n=-N}^{N} |C_{N,n}|^{2}, \quad \sum_{N=0}^{\infty}p_{N} = 1.
	\end{equation}
	The mean photon number in the interferometer is
	\begin{equation}
		\label{eq:meanphoton1}
		\langle \Psi | \ada + \bdb |\Psi \rangle = \sum_{N=0}^{\infty} p_{N} N,
	\end{equation}
	We can now compute
	\begin{eqnarray}
		\langle \Psi | H | \Psi \rangle & = & \frac{1}{2} \sum_{N=0}^{\infty} \sum_{n=-N}^{N} |C_{N,n}|^{2} n , \nonumber \\
		\langle \Psi | H^{2} | \Psi \rangle & = & \frac{1}{4} \sum_{N=0}^{\infty}  \sum_{n=-N}^{N} |C_{N,n}|^{2} n^2 , \nonumber \\
		(\Delta H)^2  & = & \frac{1}{4} \sum_{N=0}^{\infty}  \sum_{n=-N}^{N} |C_{N,n}|^{2} n^2  - \bigg[ \frac{1}{2} \sum_{N=0}^{\infty}  \sum_{n=-N}^{N} |C_{N,n}|^{2} n  \bigg]^{2}.
	\end{eqnarray}
	Fixing the mean photon number constrains only $p_{N}$ as can be seen from Eq.(\ref{eq:meanphoton1}). The choice of $C_{N,n}$, for each $N$, that maximizes $\langle \Delta^{2} H \rangle$ is
	\begin{equation}
		\label{eq:maxq}
		|C_{N,-N}|^{2}=|C_{N,N}|^{2}=\frac{p_{N}}{2}, 
	\end{equation}
	with $ C_{N,n} = 0 \; {\rm for} \; n=-(N-1), \ldots N-1$. For this choice,
	\begin{equation}
		\label{eq:variance1}
		(\Delta H)^2  = \frac{1}{4} \sum_{N=0}^{\infty}p_{N} N^{2}.
	\end{equation}
	By setting $|p_{0}|^{2} = 1-\bar{N}/N$ and $|p_{N}|^{2} = \bar{N}/N$ for $N=\lceil \kappa/ \bar{N}\rceil$, we can make
	$ \Delta H$ bigger than an arbitrary constant $\kappa$ while simultaneously keeping the mean photon number fixed. So we see that the mean photon number of the input state does not constrain the minimum achievable measurement uncertainty. We therefore need an additional constraint to use the scaling of $\Delta \varphi$ with $\bar{N}$ as a meaningful measure of the performance of an interferometer. For the states we discuss in this work, the additional constraint is that they are product states. 
	%%%%%%%%%%%%%%%%%%%%%%%%%%%%%%%%%%%%%%%%%%%%%%%%%%%%%%%%%%%%%%%%%%%%%%%%%%%%%%%%
	
	\section{Two-Mode Gaussian Product States with Losses \label{sec:gaussian}}
	
	A general two-mode pure Gaussian product state has the form
	\begin{equation}
		\label{eq:input}
		|\psi_{in}\rangle = R_{a}(\omega_a) D_a(\alpha_a) S_a(\xi_a) |0\rangle \otimes R_b(\omega_b) D_b(\alpha_b) S_b(\xi_b) |0\rangle 
	\end{equation}
	where $R_a(\omega_a)=e^{i a^\dagger a \omega_a}$, $D_a(\alpha_a)=e^{\alpha_a a^\dagger-\alpha_a^* a}$, and $S_a(\xi_a)=e^{\frac{1}{2}({\xi_a}^* a^2-\xi_a {a^\dagger}^2)}$ are the rotation, displacement, and squeezing operators respectively for the mode labelled $a$ with $\alpha_a=|\alpha_a| e^{i \beta_a}$, $\xi_a=r_a e^{i\theta_a}$ and similarly for the mode labeled $b$.  The input two mode product state of an interferometer like the one in Fig.~(\ref{fig1}) goes first through a beam splitter described by the operation,
	\[ B = e^{-i \frac{\pi}{4} (a^{\dagger}b^{\vphantom{dagger}} + a^{\vphantom{dagger}} b)}, \]
	following which the phase to be  detected is imprinted on the state through a pair of symmetric phase shifts described by $R_{a}(\varphi/2)$ and $R_{b}(-\varphi/2)$ in the ideal case.  
	
	To model the photon losses in a real inferometer, we introduce additional hypothetical beam splitters in each of the arms after the first beam splitter as shown in Fig.\ref{fig1}. We trace out the reflected part of the beam that account for the photon losses. The Kraus operators, $\mathcal{K}_{a,p}$, describing the photon losses for the mode $a$ are~\cite{dorner2009optimal}
	\begin{eqnarray}
		\label{kraus}
		\mathcal{K}_{a,p} &=& (1-\eta_a)^{p/2} \eta_a^{\frac{a\dagger a}{2}} \frac{a^p}{\sqrt{p!}}; \qquad \sum_{p=0}^{\infty}\mathcal{K}_{a,p} {\mathcal{K}_{a,p}}^\dagger =\mathbbm{1},
	\end{eqnarray}
	where $\eta_{a}$ is the transmissivity of the beam splitter and similarly for mode $b$ also. The state at the output end of the interferometer is given by,
	\begin{eqnarray}
		\label{frho}
		\rho^{\rm f}&=&\sum_{p,q} U_{ab}\mathcal{K}_{a,p}\mathcal{K}_{b,q} B \rho^{\rm in} B^\dagger {\mathcal{K}^{\dagger}_{a,p}}{\mathcal{K}^{\dagger}_{b,q}} U_{ab}^{\dagger},
	\end{eqnarray}
	where $\rho^{\rm in}=|\psi_{\rm in}\rangle\langle \psi_{\rm in}|$ and $U_{ab}=R_{a}(\varphi/2)R_{b}(-\varphi/2)$ is the unitary that imprints the phase onto the state. Note that the unitary phase operator commutes with the photon loss operators and hence the ordering of the two does not matter. For finding the Fisher information we compute the following expectation values with respect to $\rho^{\rm f}$:
	\begin{eqnarray}
		\label{momentsgaussa}
		\langle a\rangle &= &\frac{1}{\sqrt{2}}e^{i \varphi/2} \sqrt{\eta_{a} } \Big[ e^{i \left(\beta _a+\omega _a\right)} |\alpha_a| -i e^{i \left(\beta _b+\omega _b\right)} |\alpha_b| \Big] , \nonumber \\
		\langle a^2 \rangle &= & \frac{1}{2} e^{ i \varphi} \eta_{a}   \Big\{ \Big[ e^{i (\beta _a+\omega _a)} |\alpha_a| -i e^{i (\beta _b+\omega _b)} |\alpha_b| \Big]^2 \nonumber \\
		&& \qquad  -    e^{i (\theta _a +2\omega _a)} \cosh(r_a) \sinh(r_a) + e^{i (\theta _b+2\omega _b)} \cosh (r_b) \sinh (r_b) \Big\}\nonumber \\
		\langle a^\dagger a\rangle &= & \frac{1}{2} \eta_{a}  \big[ |\alpha_a| ^2+|\alpha_b|^2 +\sinh^{2}(r_a) +\sinh^{2}(r_b) \nonumber \\
		&& \qquad \qquad  -2 |\alpha_a|  |\alpha_b| \sin(\beta _a-\beta _b+\omega _a-\omega _b ) \big],
	\end{eqnarray}
	\begin{eqnarray}
		\label{momentsgaussb}
		\langle b\rangle &= &\frac{1}{\sqrt{2}}e^{-i \varphi/2} \sqrt{\eta_{b}} \Big[e^{i (\beta _b+\omega_b)} |\alpha_b| -i e^{i (\beta _a+\omega _a)} |\alpha_a|  \Big] , \nonumber \\
		\langle b^2 \rangle &= & \frac{1}{2} e^{- i\varphi} \eta_{b}  \Big\{ \Big[ e^{i (\beta _b+\omega _b)} |\alpha_b|  -i e^{i (\beta _a+\omega _a)} |\alpha_a| \Big]^2 \nonumber \\
		&& \qquad + e^{i (\theta _a+2\omega _a)} \cosh(r_a) \sinh(r_a) -  e^{i (\theta _b+2\omega _b)}  \cosh(r_b) \sinh(r_b) \Big\},\nonumber \\
		\langle b^\dagger b\rangle &=&\frac{1}{2} \eta_{b}  \big[ |\alpha_a| ^2+|\alpha_b|^2 +\sinh^{2}(r_a) +\sinh^{2}(r_b) \nonumber \\
		&& \qquad \qquad  +2 |\alpha_a|  |\alpha_b| \sin(\beta _a-\beta _b+\omega _a-\omega _b) \big].
	\end{eqnarray}
	and
	\begin{eqnarray}
		\label{momentsgaussab}
		\langle ab \rangle & = & -\frac{1}{2} i \sqrt{\eta_{a} \eta_{b}} \big[ e^{2i(\beta{a} + \omega_{a})} |\alpha_{a}|^{2}+ e^{2i(\beta_{b} + \omega_{b})}|\alpha_{b}|^{2}  \nonumber \\
		&& \qquad -  e^{i (\theta _a +2\omega _a)} \cosh(r_a) \sinh(r_a) - e^{i (\theta _b+2\omega _b)} \cosh (r_b) \sinh (r_b) \big] \nonumber \\
		\langle ab^{\dagger} \rangle & = & \frac{1}{2} i e^{i\varphi} \sqrt{\eta_{a} \eta_{b}} \big[ |\alpha_a|^2 - |\alpha_b|^2 +\sinh^{2}(r_a) -\sinh^{2}(r_b) \nonumber \\
		&& \qquad \qquad  -2i |\alpha_a|  |\alpha_b| \cos(\beta _a-\beta _b+\omega _a-\omega _b) \big].
	\end{eqnarray}
	
	In fact we first calculate the expectation values above without considering the photon losses, then we include the photon losses using the transformation $\langle a \rangle\rightarrow\sqrt{\eta_{a}}\langle a \rangle$, $\langle a^2\rangle\rightarrow\eta_{a} \langle a^2\rangle$, and $\langle a^\dagger a \rangle\rightarrow\eta_{a} \langle a^\dagger a \rangle$, $\langle ab \rangle \rightarrow \sqrt{\eta_{a} \eta_{b}} \langle ab \rangle $ and similarly for mode $b$. The remaining expectation values like $\langle a^{\dagger} \rangle$, $\langle a^{\dagger^{2}} \rangle$ etc.~that are required to compute the covariance matrix $\Sigma$ can be obtained from the above by complex conjugation as appropriate. 
	
	Using Eq.~(\ref{eq:covar}) and assuming that the photon losses in either arm is the same, $\eta_{a} = \eta_{b} = \eta$ we obtain
	\begin{equation}
		\Sigma = \frac{1}{2}\left(
		\begin{array}{cccc}
			P & Q & -e^{i\varphi}R& S \\
			Q^{*} & P & S &e^{-i\varphi} R \\
			-e^{-i\varphi}R^{*} & S^{*} & P & Q^{*} \\
			S^{*} &e^{i\varphi} R^{*} & Q & P \\
		\end{array}
		\right) ,
	\end{equation}
	where
	\begin{eqnarray*}
		\label{eq:abvals}
		P & = & 2(1 - \eta) + \eta \big[ \cosh (2r_{a}) + \cosh (2r_{b}) \big], \nonumber \\
		Q & = & i \eta e^{i\varphi} \big[ \cosh(2 r_{a}) - \cosh(2r_{b}) \big] ,\nonumber \\
		R & = & \eta \big[ e^{i (\theta_{a}+ 2 \omega_{a})} \sinh (2r_{a}) - e^{i (\theta_{b}+2\omega_{b})} \sinh  (2r_{b})  \big],  \nonumber  \\
		S & = & i \eta \big[ e^{i (\theta_{a}+ 2 \omega_{a})} \sinh (2r_{a}) + e^{i (\theta_{b}+2\omega_{b})} \sinh  (2r_{b})  \big].
	\end{eqnarray*}
	The symplectic eigenvalues of $A=K\Sigma$ are given by
	\begin{eqnarray}
		\label{eq:symplectic2}
		\tau_{1} & = &  \sqrt{ 1 + 2 \eta (1- \eta) (\cosh (2r_{a})-1) }, \nonumber \\
		\tau_{2} & = &  \sqrt{ 1 + 2 \eta (1-\eta) (\cosh (2r_{b})-1) },
	\end{eqnarray}
	both of which are $\varphi$ independent and positive. Since $\dot{\tau}_{1} = \dot{\tau}_{2} = 0$, the quantum Fisher information has the form
	\[ \mathcal{I} = \mathcal{I}_{1} + \mathcal{I}_{3}, \]
	where
	\[ \mathcal{I}_{1} = \frac{1}{2(|A|-1)}  \bigg\{ |A| {\rm tr} \Big[ \big(A^{-1}\dot{A}\big)^2\Big]+ \sqrt{|\mathbbm{1}+A^2|}  {\rm tr} \Big[\big( ( \openone +A^2)^{-1}\dot{A}\big)^2\Big] \bigg\}, \]
	and
	\[ \mathcal{I}_{3} = 2 \dot{\vec{d}}^\dagger \Sigma^{-1} \dot{\vec{d}}.\]
	After a long but straightforward calculation we find that ${\mathcal I}_{1}$ has the form
	\[ {\mathcal I}_{1} = f_{1}(\eta, r_{a}, r_{b}) - g_{1}(\eta, r_{a} ,r_{b})\cos (\theta_{a} - \theta_{b}+ 2\omega_{a} - 2\omega_{b}), \]
	where $f_{1}$ and $g_{1}$ are positive valued functions. Similarly we have
	\begin{eqnarray*} 
		{\mathcal I}_{3} & = & f_{2} (\eta, |\alpha_{a}|, |\alpha_{b}|, r_{a}, r_{b}) + \eta^{2} \bigg[ \frac{|\alpha_{a}|^{2} \sinh (2 r_{b})}{\tau_{2}} \cos(\theta_{b} -2(\beta_{a} + \omega_{a} - \omega_{b}))  \\
		&& \qquad + \frac{|\alpha_{b}|^{2} \sinh (2 r_{a})}{\tau_{1}} \cos(\theta_{a} -2(\beta_{b} - \omega_{a} + \omega_{b})) \bigg]. 
	\end{eqnarray*}
	We can maximise both ${\mathcal I}_{1}$ and ${\mathcal I}_{2}$ by choosing the phases $\omega_{a}$, $\omega_{b}$, $\theta_{a}$, $\theta_{b}$, $\beta_{a}$ and $\beta_{b}$ such that $\cos (\theta_{a} - \theta_{b}+ 2\omega_{a} - 2\omega_{b})=-1$, $\cos(\theta_{b} -2(\beta_{a} + \omega_{a} - \omega_{b}))=1$ and $\cos(\theta_{a} -2(\beta_{b} - \omega_{a} + \omega_{b}))=1$. For such choice of phases, we find that:
	\begin{eqnarray}
		\label{eq:fgauss}
		\mathcal{I} &= &\eta \bigg\{ \frac{ \big[ \eta  \cosh(2r_{a}) +  \eta \cosh(2r_{b})  + (1-\eta) \cosh(2r_{a} - 2r_{b}) - \eta -1 \big]\sinh^{2}(r_{a} + r_{b})}{(1-2\eta + 2\eta^{2}) [ \cosh(2r_{a}) +  \cosh(2r_{b})  ] + 2\eta(1-\eta)\cosh(2r_{a}) \cosh(2r_{b}) -2 [1-\eta(1-\eta)]} \nonumber \\
		& & \qquad \qquad +\frac{ e^{r_a+r_b} \left(|\alpha _a|^2+|\alpha _b|^2\right)-2 \eta  \big[ e^{r_b} \sinh(r_a) |\alpha _a|^2+e^{r_a} \sinh(r_b) |\alpha _b|^2 \big] }{\big[ \cosh(r_{a}) + (1-2\eta)\sinh(r_{a}) \big]\big[ \cosh(r_{b}) + (1-2\eta)\sinh(r_{b}) \big]} \bigg\}.
	\end{eqnarray}
	Note that the quantum Fisher information we obtain is independent of $\varphi$. The remaining phases $\omega_{a}$, $\omega_{b}$, $\theta_{a}$, $\theta_{b}$, $\beta_{a}$ and $\beta_{b}$ are removed while maximising ${\mathcal I}$.
	
	\section{Special cases \label{sec:special}}
	
	The input product state in Eq.~(\ref{eq:input}) has mean photon number
	\begin{equation}
		\label{eq:meanphoton}
		\bar{N} = |\alpha_{a}|^{2} + \sinh^{2}(r_{a}) + |\alpha_{b}|^{2} +\sinh^{2}(r_{b}). 
	\end{equation}
	In the following, for various examples we compare the quantum Fisher information computed using Eq.~(\ref{eq:fgauss}) with best possible scaling of the measurement precision given using classical (coherent) states as input to an interferometer with losses, namely $1/\sqrt{\eta\bar{N}}$. Specifically we will be focusing on the ratio ${\mathcal J} \equiv \sqrt{{\mathcal I}/\eta\bar{N}}$. Measurement precision that is better than the classical case will correspond to the ratio growing as a function of $\bar{N}$, while saturation of the ratio would correspond to improvements over classical scaling by a constant factor only. 
	
	\subsection{Dual Squeezed Vacuum State}
	In~\cite{lang2014optimal} the dual squeezed vacuum (DSV) state is identified as the optimal product input state with fixed mean photon number into an ideal optical interferometric set-up. The DSV state is,
	\begin{eqnarray}\label{rhoin}
		\rho^{\rm in}= |r\rangle\langle r| \otimes |-r\rangle \langle -r|,
	\end{eqnarray}  
	where $|r\rangle=S(r)|0\rangle$ with $r$ real. The DSV state is obtained as a special case of the state in  Eq.~(\ref{eq:input}) by setting $|\alpha_{a}|=|\alpha_{b}|=\beta_{a} = \beta_{b}=\omega_{a}=\omega_{b}=0$, $r_{a} = r_{b} = r$ and $\theta_{a}=0,\theta_{b}=\pi$. Substituting these values into Eq.~(\ref{eq:fgauss}) with $\eta=1$, we recover the result from~\cite{lang2014optimal} that the phase measurement scales as, 
	\begin{equation}
		\label{eq:dsvfisher1}
		\Delta \varphi \geq \frac{1}{\sqrt{\bar{N}(\bar{N}+2)}}, \qquad \bar{N} = 2 \sinh^{2} (r).
	\end{equation}
	Note that since $|\alpha_{a}|$ and $|\alpha_{b}|$ are zero, maximising the Fisher information involves only setting $\cos(\theta_{a} - \theta_{b}+2\omega_{a}-2\omega_{b})=-1$ which is satisfied for the DSV state. 
	
\begin{figure}[!htb]
		\centering
		\resizebox{8.5cm}{5.5 cm}{\includegraphics{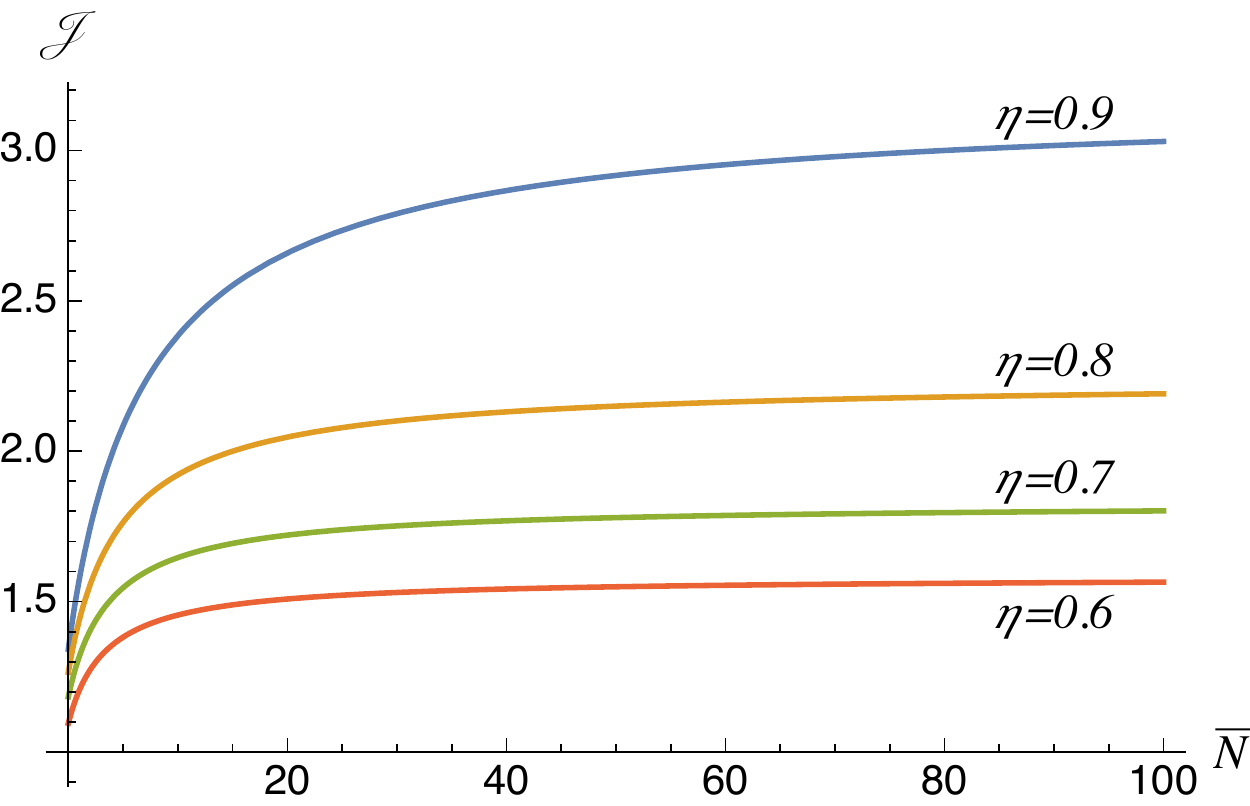}}
		\caption{The ratio ${\mathcal J}$ for the DSV state is plotted against the mean photon number for different values of transmissivities ($\eta$). For finite photon losses, the DSV state does not give any improvement in scaling of the measurement uncertainty with respect to the mean photon number. Improvement by a constant factor is however possible. \label{fig2}}
	\end{figure}

	In the presence of photon losses, the quantum Fisher information for the DSV state is
	\begin{eqnarray}\label{eq:I3}
		\mathcal{I}_{\rm DSV}&=&\frac{\eta^2 \bar{N}(\bar{N}+2)}{1+ (1-\eta)\eta \bar{N}}.
	\end{eqnarray}
	The corresponding uncertainty in the phase measurement is
	\begin{equation}\label{dphi}
		{\Delta \varphi} \;\geq \; \frac{1}{\sqrt{\mathcal{I}}}\;=\;\sqrt{\frac{1+ (1-\eta)\eta \bar{N}}{\eta^2 \bar{N}(\bar{N}+2)}}.
	\end{equation}
	and
	\begin{equation}\label{eq:ratio}
		{\mathcal J}_{\rm DSV} = \sqrt{\frac{\mathcal{I}_{\rm DSV}}{\eta\bar{N}}} = \frac{\Delta\varphi_{\rm classical}}{\Delta \varphi_{\rm DSV}}=\sqrt{\frac{\eta(\bar{N}+2)}{1+ (1-\eta)\eta \bar{N}}}.
	\end{equation}
	In the limit of large $\bar{N}$ (large squeezing) we find that for $\eta \neq1$, ${\mathcal J}$ saturates to $\sqrt{1/(1-\eta)}$ indicating that with photon losses the DSV states in an interferometer will not outperform the classical case in the scaling for the measurement uncertainty. The same is already known to be true in the case of quantum metrology using states with fixed particle number~\cite{DemkowiczDobrzanski:2012gl,DemkowiczDobrzanski:2013fs,Shaji:2007bt}. However, as shown in~\cite{lang2014optimal}, in the ideal case where $\eta=1$, ${\mathcal J}$ grows with ${\bar N}$ as $\sqrt{\bar{N}+2}$, giving Heisenberg limited scaling. The improvement through a constant factor in the presence of photon loss is shown in Fig.~(\ref{fig2}) where ${\mathcal J}$ is plotted as a function of $\bar{N}$ for different values of $\eta$. 
	
	\subsection{Dual Squeezed, Displaced Vacuum State}
	
	The dual squeezed, displaced vacuum (DSDV) state is generated by symmetrically displacing and squeezing both modes so that $|\alpha_{a}|=|\alpha_{b}|=\alpha$, $r_{a} = r_{b} = r$. One choice of phases that maximise the quantum Fisher information is $\omega_{a} = \omega_{b} = \theta_{a} = \beta_{b} =0$, $\theta_{b} = 2\beta_{a} = \pi$. The DSDV state can be labeled as $|\psi_{\rm in} \rangle = |i\alpha, r\rangle \otimes |\alpha, -r\rangle$. 
	Using these parameters in Eq.~(\ref{eq:fgauss}) we obtain
	\begin{equation}
		\label{eq:fisherDSDV}
		\mathcal{I}_{\rm DSDV} =  \eta \bigg\{ \frac{  2e^{2 r} \alpha^{2}}{\eta + e^{2 r} (1-\eta ) } +\frac{\eta \sinh^2(2r)}{1+\eta(1-\eta) (\cosh(2r)-1)} \bigg\}.
	\end{equation}
	For the DSDV state, $\bar{N}=2\left( \alpha^2+\sinh^2(r)\right)$ and
	\begin{equation}
		{\mathcal J}_{DSDV} = \sqrt{\frac{1}{2( \alpha^2+\sinh^2(r))} \bigg\{ \frac{  2e^{2 r} \alpha^{2}}{\eta + e^{2 r} (1-\eta ) } +\frac{\eta \sinh^2(2r)}{1+\eta(1-\eta) (\cosh(2r)-1)} \bigg\}}. 
	\end{equation}
	
	There are two types of large $\bar{N}$ limits we can take for the DSDV states. The first is a classical limit wherein we assume that the displacement $\alpha$ is large with negligible or no squeezing ($r=0$). In this case we have ${\mathcal J}_{DSDV}^{(c)} = 1$. With or without photon losses, the DSDV states offer no advantage over the classical case as expected. The quantum limit corresponds to $\alpha=0$; in which case the DSDV state reduces to the DSV state discussed previously.

	\begin{figure}[!htb]
		\centering
		\resizebox{8.5cm}{5.5cm}{\includegraphics{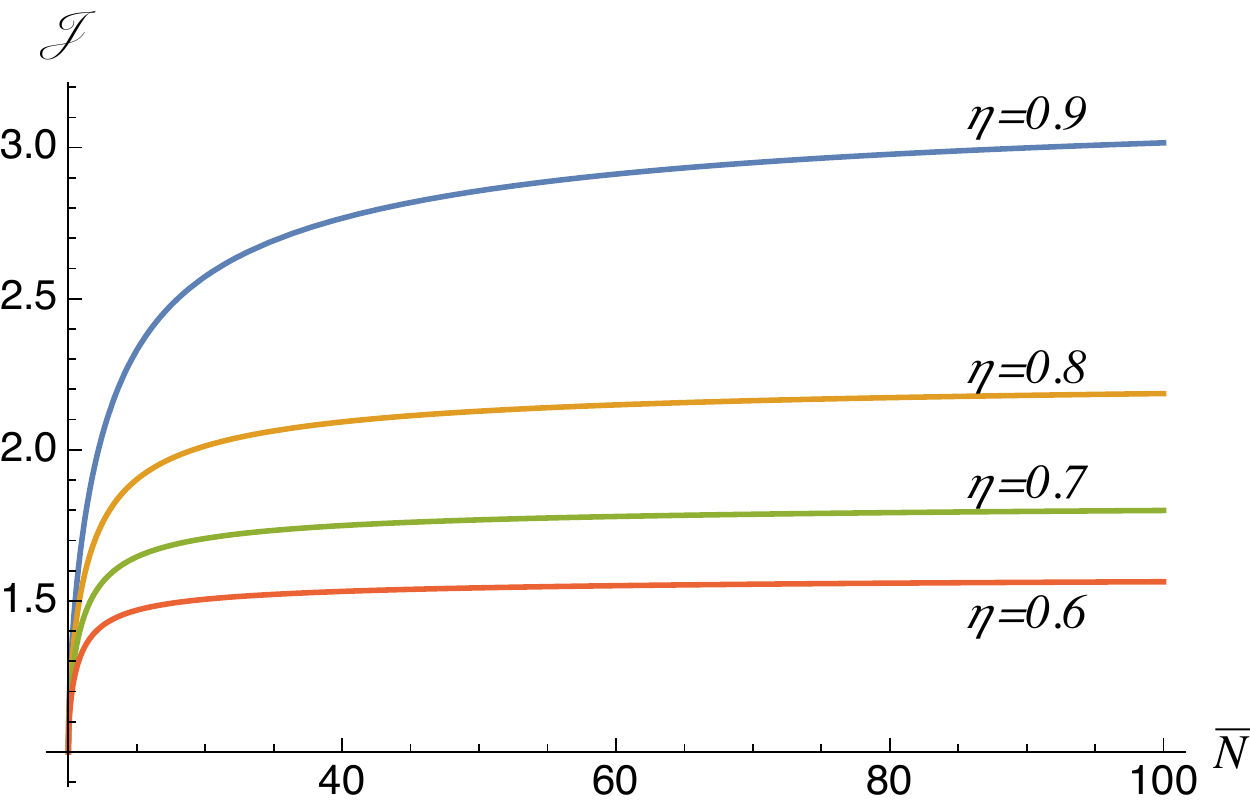}}
		\caption{The ratio ${\mathcal J}$ is plotted against $\bar{N} = 2(\alpha^{2} + \sinh^{2}(r))$ for the DSDV state in a lossy interferometer with different values of $\eta$ for which $\alpha^{2} =10$ and the squeezing $r$ is varied. We see that the graph is almost identical to that for the DSV state in Fig.~\ref{fig2} indicating that the displacement has minimal effect on modulating the limits on $\delta \varphi$.  \label{fig3}}
	\end{figure}
	
	\begin{figure}[!htb]
		\centering
		\resizebox{8.5cm}{5.5cm}{\includegraphics{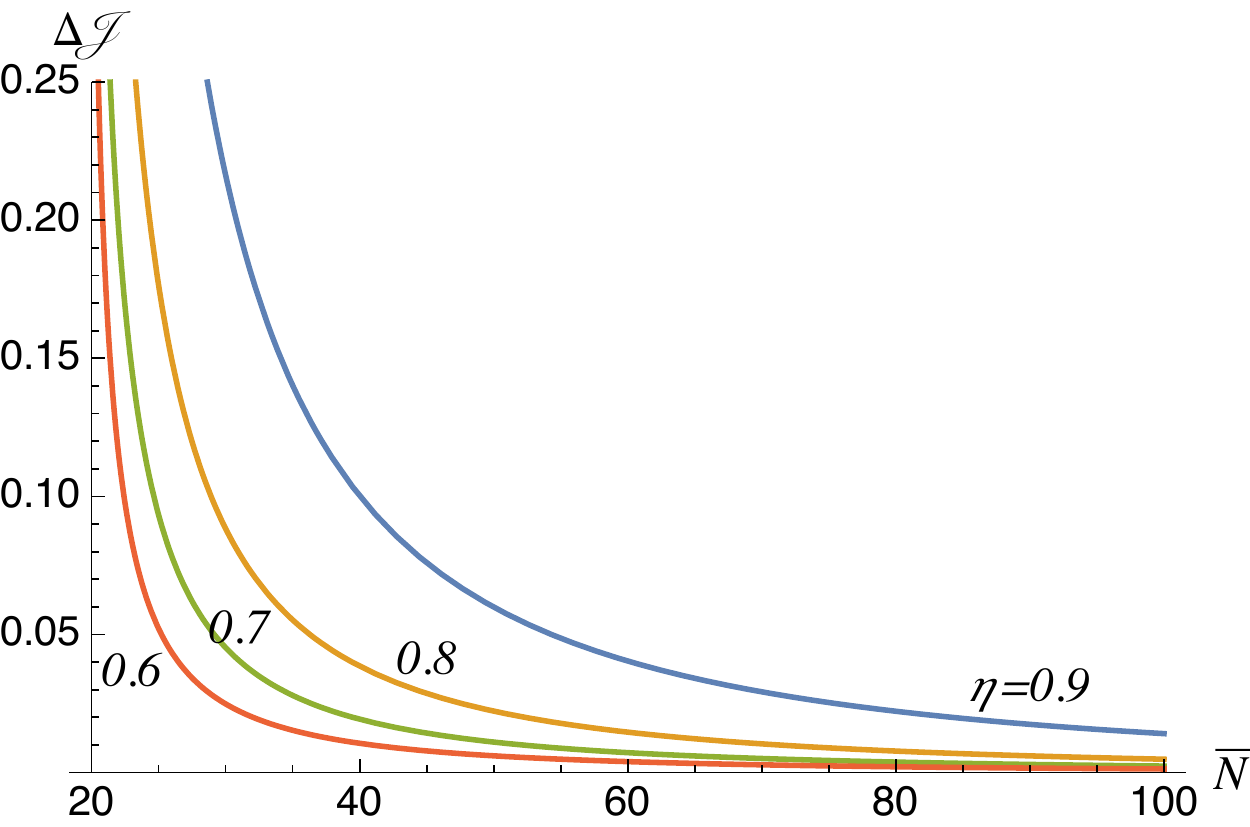}}
		\caption{The difference $\Delta {\mathcal J} ={\mathcal J}_{\rm DSDV} -  {\mathcal J}_{\rm DSV}$ with $\alpha^{2}$ fixed at 10 for the DSDV state is plotted as a function of $\bar{N}$ highlighting the difference in measurement precision that the displacement operator brings about.\label{fig4}}
	\end{figure}
	
	In Fig.~\ref{fig3}, the ratio ${\mathcal J}$ is plotted for a DSDV state for which the contribution to the mean photon number due to the displacement operator is kept constant at $\alpha^{2} = 10$ and the mean photon number is varied by changing the squeezing. As with the DSV state, we find that the improvement in the measurement precision over the classical case is through a constant factor only, and the saturation value of the improvement is controlled by the photon loss parametrised by $\eta$. We also note that the graphs in Fig.~\ref{fig3} are very similar to those for the DSV state, shown in Fig.~\ref{fig2} and both saturate to the same value for a given $\eta$. This indicates that the displacement has little or no role in improving the measurement precision beyond the classical case as expected. In Fig.~\ref{fig4} we plot the difference between ${\mathcal J}_{\rm DSDV}$ and ${\mathcal J}_{\rm DSV}$ as a function of the mean photon number and we see that the difference becomes very small as $\bar{N}$ increases. 
	
	\begin{figure}[!htb]
		\centering
		\resizebox{8.5cm}{5.5cm}{\includegraphics{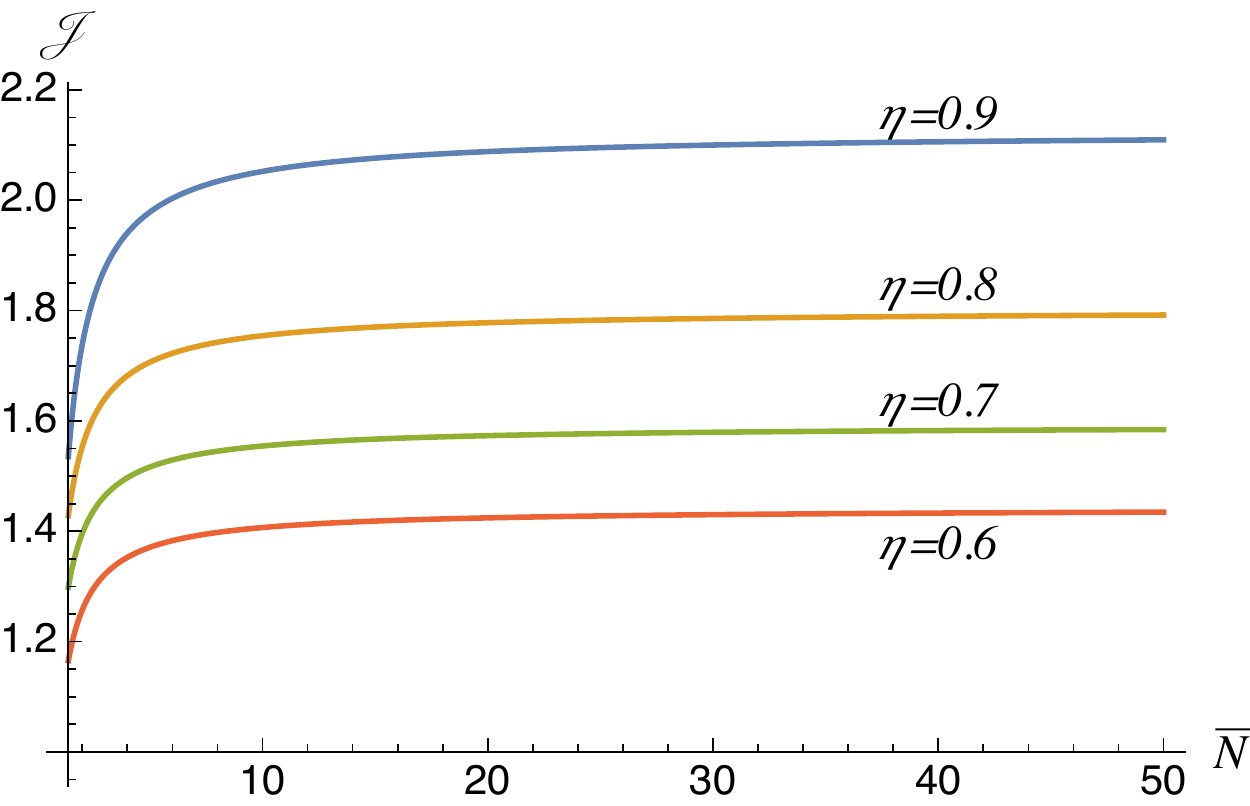}}
		\caption{The ratio ${\mathcal J}$ is plotted against $\bar{N} = 2(\alpha^{2} + \sinh^{2}(r))$ for the DSDV state in a lossy interferometer with different values of $\eta$ for which the squeezing parameter $r =1$ and the displacement $\alpha$ is varied. The maximum value of the improvement by a constant factor over the classical state that is obtained by the DSDV state is attained quickly for relatively low $\bar{N}$, with no further substantial improvements as a function of the displacement. \label{fig5}}
	\end{figure}
	
	In Fig.~\ref{fig5} we plot ${\mathcal J}_{\rm DSDV}$ for fixed squeezing ($r=1$) with $\bar{N}$ varied by changing $\alpha$. We see that the saturation values of the improvement are controlled by $r$ and $\eta$ and the graphs already reach their respective maximum values for relatively small displacements again highlighting that the displacement in the DSDV states has very little role to play in providing an improvement in the achievable interferometric measurement precision over the classical case. 
	
\begin{figure}[!htb]
		\centering
		\resizebox{9.5cm}{8cm}{\includegraphics{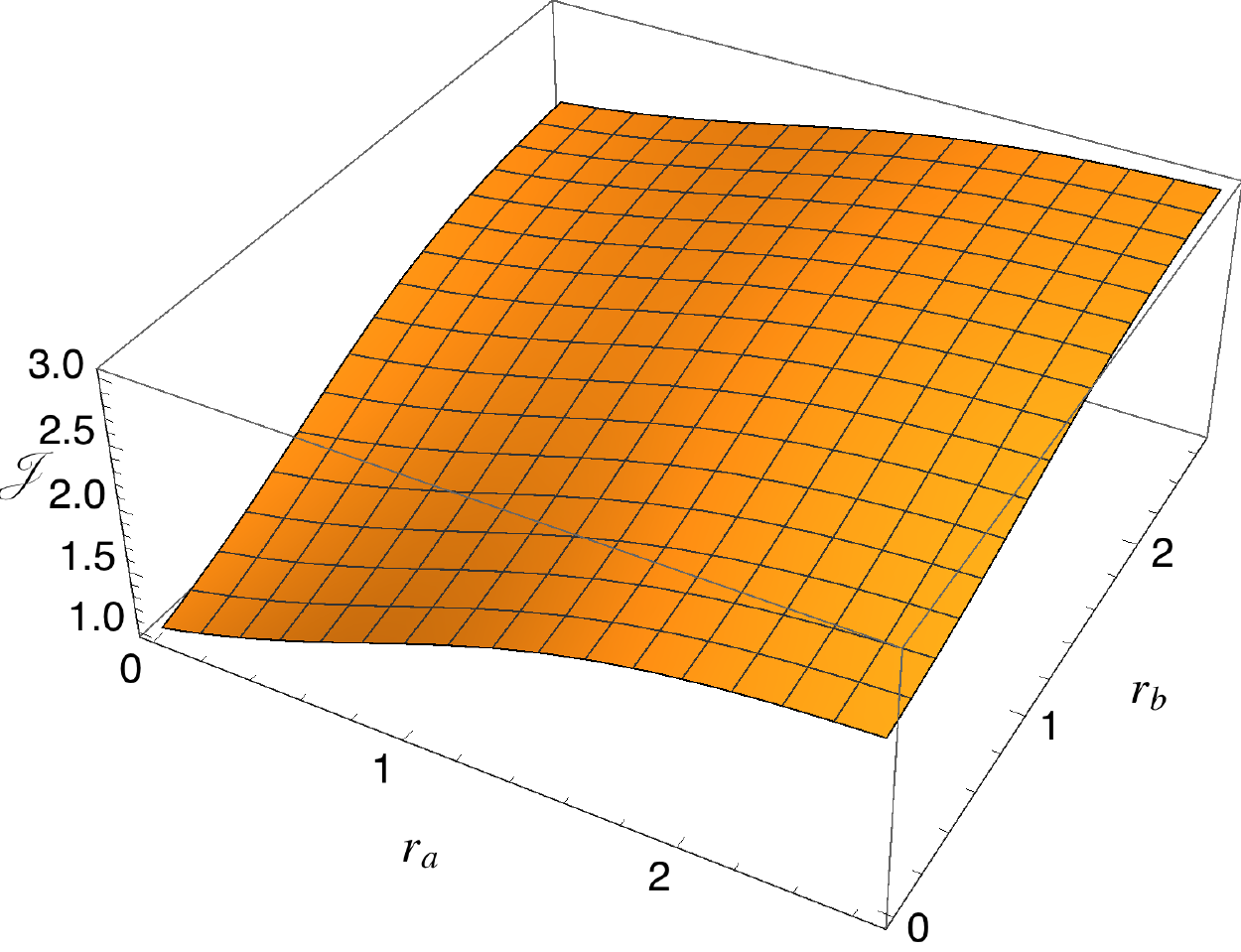}}
		\caption{The ratio ${\mathcal J}$ is plotted as a function of the squeezing $r_{a}$ and $r_{b}$ of each of the two modes for a DSDV state with a fixed mean photon number, $\bar{N}=100$ in an interferometer with $\eta=0.9$. The mean photon number is kept fixed by choosing $|\alpha_{j}|^{2} = \bar{N}/2 - \sinh^{2} (r_{j}), \; j=a,b$.\label{fig6}}
	\end{figure}

	To explore the effect of the trade-off between displacement and squeezing in the DSDV states, In Fig.~\ref{fig6}, we plot the ratio ${\mathcal J}$ as a function of the squeezing in each mode as measured by the value of $r_{a}$ and $r_{b}$ for a state with fixed mean photon number $\bar{N}=100$ in an interferometer with $\eta=0.9$. We assign a mean photon number of $\bar{N}/2$ for each mode by choosing $|\alpha_{j}|^{2} = \bar{N}/2 - \sinh^{2} (r_{j}), \; j=a,b$. Again it is clear that any advantage over the classical case is due to the squeezing in either mode and that in the presence of photon loss no advantage in the scaling of the measurement uncertainty is forthcoming 
	
	%%%%%%%%%%%%%%%%%%%%%%%%%%%%%%%%%%%%%%%%%%%%%%%%%%%%%%%%%%%%%%%%%%%%%%%%%%%%%%%%%%%
	\section{Summary}
	
	We have presented a systematic exploration of the performance of two-mode Gaussian product states in an interferometer beset by photon losses. The choice of these states was motivated by the observation in~\cite{lang2014optimal} that such states are optimal in ideal interferometers among input product states with fixed mean photon number.  We obtained an expression for the quantum Fisher information in the presence of photon losses for such states in a linear, symmetric interferometer. We proceeded to use this expression to find the quantum Cramer-Rao bound on the minimum uncertainties in phase sensing when using DSV and DSDV states in a lossy interferometric set-up. We have shown that both states do not attain Heisenberg-limited scaling for the measurement uncertainty if there are photon losses present, but it still performs better than classical states of light albeit through constant factors. Our results reduce to those in~\cite{lang2014optimal} in the absence of losses. 
	
	We specifically studied the ratio between measurement uncertainty when using classical-like (coherent) states in a lossy interferometer and the uncertainty using the states that we consider.  We find that as anticipated in~\cite{DemkowiczDobrzanski:2012gl,DemkowiczDobrzanski:2013fs,Shaji:2007bt} for the case of fixed photon number states, for any finite photon loss, for fixed mean photon number states of the kind we considered also the ratio does saturate and no improvement in the scaling of the measurement precision is forthcoming even with squeezing of the input modes. We find that improvements by substantial, but constant factors are however possible. 
	
	\acknowledgments
	This work is supported in part by the Ramanujan Fellowship programme (No.~SR/S2/RJN-01/2009) of the Department of Science and Technology, Government of India. 
	
	\bibliography{int} 

%merlin.mbs apsrev4-1.bst 2010-07-25 4.21a (PWD, AO, DPC) hacked
%Control: key (0)
%Control: author (8) initials jnrlst
%Control: editor formatted (1) identically to author
%Control: production of article title (-1) disabled
%Control: page (0) single
%Control: year (1) truncated
%Control: production of eprint (0) enabled
\begin{thebibliography}{30}%
\makeatletter
\providecommand \@ifxundefined [1]{%
 \@ifx{#1\undefined}
}%
\providecommand \@ifnum [1]{%
 \ifnum #1\expandafter \@firstoftwo
 \else \expandafter \@secondoftwo
 \fi
}%
\providecommand \@ifx [1]{%
 \ifx #1\expandafter \@firstoftwo
 \else \expandafter \@secondoftwo
 \fi
}%
\providecommand \natexlab [1]{#1}%
\providecommand \enquote  [1]{``#1''}%
\providecommand \bibnamefont  [1]{#1}%
\providecommand \bibfnamefont [1]{#1}%
\providecommand \citenamefont [1]{#1}%
\providecommand \href@noop [0]{\@secondoftwo}%
\providecommand \href [0]{\begingroup \@sanitize@url \@href}%
\providecommand \@href[1]{\@@startlink{#1}\@@href}%
\providecommand \@@href[1]{\endgroup#1\@@endlink}%
\providecommand \@sanitize@url [0]{\catcode `\\12\catcode `\$12\catcode
  `\&12\catcode `\#12\catcode `\^12\catcode `\_12\catcode `\%12\relax}%
\providecommand \@@startlink[1]{}%
\providecommand \@@endlink[0]{}%
\providecommand \url  [0]{\begingroup\@sanitize@url \@url }%
\providecommand \@url [1]{\endgroup\@href {#1}{\urlprefix }}%
\providecommand \urlprefix  [0]{URL }%
\providecommand \Eprint [0]{\href }%
\providecommand \doibase [0]{http://dx.doi.org/}%
\providecommand \selectlanguage [0]{\@gobble}%
\providecommand \bibinfo  [0]{\@secondoftwo}%
\providecommand \bibfield  [0]{\@secondoftwo}%
\providecommand \translation [1]{[#1]}%
\providecommand \BibitemOpen [0]{}%
\providecommand \bibitemStop [0]{}%
\providecommand \bibitemNoStop [0]{.\EOS\space}%
\providecommand \EOS [0]{\spacefactor3000\relax}%
\providecommand \BibitemShut  [1]{\csname bibitem#1\endcsname}%
\let\auto@bib@innerbib\@empty
%</preamble>
\bibitem [{\citenamefont {Michelson}\ and\ \citenamefont
  {Morley}(1887)}]{Michelson:1887ka}%
  \BibitemOpen
  \bibfield  {author} {\bibinfo {author} {\bibfnamefont {A.~A.}\ \bibnamefont
  {Michelson}}\ and\ \bibinfo {author} {\bibfnamefont {E.~W.}\ \bibnamefont
  {Morley}},\ }\href@noop {} {\bibfield  {journal} {\bibinfo  {journal}
  {American Journal of Science}\ }\textbf {\bibinfo {volume} {s3-34}},\
  \bibinfo {pages} {333} (\bibinfo {year} {1887})}\BibitemShut {NoStop}%
\bibitem [{\citenamefont {Abbott}\ \emph {et~al.}(2016)\citenamefont {Abbott},
  \citenamefont {Abbott}, \citenamefont {Abbott}, \citenamefont {Abernathy},
  \citenamefont {Acernese}, \citenamefont {Ackley}, \citenamefont {Adams},
  \citenamefont {Adams}, \citenamefont {Addesso}, \citenamefont {Adhikari}
  \emph {et~al.}}]{abbott2016observation}%
  \BibitemOpen
  \bibfield  {author} {\bibinfo {author} {\bibfnamefont {B.}~\bibnamefont
  {Abbott}}, \bibinfo {author} {\bibfnamefont {R.}~\bibnamefont {Abbott}},
  \bibinfo {author} {\bibfnamefont {T.}~\bibnamefont {Abbott}}, \bibinfo
  {author} {\bibfnamefont {M.}~\bibnamefont {Abernathy}}, \bibinfo {author}
  {\bibfnamefont {F.}~\bibnamefont {Acernese}}, \bibinfo {author}
  {\bibfnamefont {K.}~\bibnamefont {Ackley}}, \bibinfo {author} {\bibfnamefont
  {C.}~\bibnamefont {Adams}}, \bibinfo {author} {\bibfnamefont
  {T.}~\bibnamefont {Adams}}, \bibinfo {author} {\bibfnamefont
  {P.}~\bibnamefont {Addesso}}, \bibinfo {author} {\bibfnamefont
  {R.}~\bibnamefont {Adhikari}},  \emph {et~al.},\ }\href {\doibase
  10.1103/PhysRevLett.116.061102} {\bibfield  {journal} {\bibinfo  {journal}
  {Phys. Rev. Lett.}\ }\textbf {\bibinfo {volume} {116}},\ \bibinfo {pages}
  {061102} (\bibinfo {year} {2016})}\BibitemShut {NoStop}%
\bibitem [{\citenamefont {Adhikari}(2014)}]{Adhikari:2014gx}%
  \BibitemOpen
  \bibfield  {author} {\bibinfo {author} {\bibfnamefont {R.~X.}\ \bibnamefont
  {Adhikari}},\ }\href {\doibase 10.1103/RevModPhys.86.121} {\bibfield
  {journal} {\bibinfo  {journal} {Rev. Mod. Phys.}\ }\textbf {\bibinfo {volume}
  {86}},\ \bibinfo {pages} {121} (\bibinfo {year} {2014})}\BibitemShut
  {NoStop}%
\bibitem [{\citenamefont {Caves}(1980)}]{Caves:1980vh}%
  \BibitemOpen
  \bibfield  {author} {\bibinfo {author} {\bibfnamefont {C.~M.}\ \bibnamefont
  {Caves}},\ }\href {http://link.aps.org/doi/10.1103/PhysRevD.23.1693}
  {\bibfield  {journal} {\bibinfo  {journal} {Physical Review D}\ }\textbf
  {\bibinfo {volume} {23}},\ \bibinfo {pages} {1693} (\bibinfo {year}
  {1980})}\BibitemShut {NoStop}%
\bibitem [{\citenamefont {Demkowicz-Dobrza{\'n}ski}\ \emph
  {et~al.}(2015)\citenamefont {Demkowicz-Dobrza{\'n}ski}, \citenamefont
  {Jarzyna},\ and\ \citenamefont {Ko{\l}ody{\'n}ski}}]{demkowicz2015chapter}%
  \BibitemOpen
  \bibfield  {author} {\bibinfo {author} {\bibfnamefont {R.}~\bibnamefont
  {Demkowicz-Dobrza{\'n}ski}}, \bibinfo {author} {\bibfnamefont
  {M.}~\bibnamefont {Jarzyna}}, \ and\ \bibinfo {author} {\bibfnamefont
  {J.}~\bibnamefont {Ko{\l}ody{\'n}ski}},\ }\href@noop {} {\bibfield  {journal}
  {\bibinfo  {journal} {Progress in Optics}\ }\textbf {\bibinfo {volume}
  {60}},\ \bibinfo {pages} {345} (\bibinfo {year} {2015})}\BibitemShut
  {NoStop}%
\bibitem [{\citenamefont {Taylor}\ and\ \citenamefont
  {Bowen}(2016)}]{Taylor20161review}%
  \BibitemOpen
  \bibfield  {author} {\bibinfo {author} {\bibfnamefont {M.~A.}\ \bibnamefont
  {Taylor}}\ and\ \bibinfo {author} {\bibfnamefont {W.~P.}\ \bibnamefont
  {Bowen}},\ }\href {\doibase http://dx.doi.org/10.1016/j.physrep.2015.12.002}
  {\bibfield  {journal} {\bibinfo  {journal} {Physics Reports}\ }\textbf
  {\bibinfo {volume} {615}},\ \bibinfo {pages} {1 } (\bibinfo {year} {2016})},\
  \bibinfo {note} {quantum metrology and its application in
  biology}\BibitemShut {NoStop}%
\bibitem [{\citenamefont {Demkowicz-Dobrzanski}\ \emph
  {et~al.}(2009)\citenamefont {Demkowicz-Dobrzanski}, \citenamefont {Dorner},
  \citenamefont {Smith}, \citenamefont {Lundeen}, \citenamefont {Wasilewski},
  \citenamefont {Banaszek},\ and\ \citenamefont
  {Walmsley}}]{DemkowiczDobrzanski:2009gl}%
  \BibitemOpen
  \bibfield  {author} {\bibinfo {author} {\bibfnamefont {R.}~\bibnamefont
  {Demkowicz-Dobrzanski}}, \bibinfo {author} {\bibfnamefont {U.}~\bibnamefont
  {Dorner}}, \bibinfo {author} {\bibfnamefont {B.~J.}\ \bibnamefont {Smith}},
  \bibinfo {author} {\bibfnamefont {J.~S.}\ \bibnamefont {Lundeen}}, \bibinfo
  {author} {\bibfnamefont {W.}~\bibnamefont {Wasilewski}}, \bibinfo {author}
  {\bibfnamefont {K.}~\bibnamefont {Banaszek}}, \ and\ \bibinfo {author}
  {\bibfnamefont {I.~A.}\ \bibnamefont {Walmsley}},\ }\href
  {http://link.aps.org/doi/10.1103/PhysRevA.80.013825} {\bibfield  {journal}
  {\bibinfo  {journal} {Physical Review A}\ }\textbf {\bibinfo {volume} {80}},\
  \bibinfo {pages} {013825} (\bibinfo {year} {2009})}\BibitemShut {NoStop}%
\bibitem [{\citenamefont {Dorner}\ \emph {et~al.}(2009)\citenamefont {Dorner},
  \citenamefont {Demkowicz-Dobrzanski}, \citenamefont {Smith}, \citenamefont
  {Lundeen}, \citenamefont {Wasilewski}, \citenamefont {Banaszek},\ and\
  \citenamefont {Walmsley}}]{dorner2009optimal}%
  \BibitemOpen
  \bibfield  {author} {\bibinfo {author} {\bibfnamefont {U.}~\bibnamefont
  {Dorner}}, \bibinfo {author} {\bibfnamefont {R.}~\bibnamefont
  {Demkowicz-Dobrzanski}}, \bibinfo {author} {\bibfnamefont {B.~J.}\
  \bibnamefont {Smith}}, \bibinfo {author} {\bibfnamefont {J.~S.}\ \bibnamefont
  {Lundeen}}, \bibinfo {author} {\bibfnamefont {W.}~\bibnamefont {Wasilewski}},
  \bibinfo {author} {\bibfnamefont {K.}~\bibnamefont {Banaszek}}, \ and\
  \bibinfo {author} {\bibfnamefont {I.~A.}\ \bibnamefont {Walmsley}},\ }\href
  {\doibase 10.1103/PhysRevLett.102.040403} {\bibfield  {journal} {\bibinfo
  {journal} {Phys. Rev. Lett.}\ }\textbf {\bibinfo {volume} {102}},\ \bibinfo
  {pages} {040403} (\bibinfo {year} {2009})}\BibitemShut {NoStop}%
\bibitem [{\citenamefont {Dowling}(2008)}]{dowling08a}%
  \BibitemOpen
  \bibfield  {author} {\bibinfo {author} {\bibfnamefont {J.~P.}\ \bibnamefont
  {Dowling}},\ }\href {http://www.informaworld.com/10.1080/00107510802091298}
  {\bibfield  {journal} {\bibinfo  {journal} {Contemporary Physics}\ }\textbf
  {\bibinfo {volume} {49}},\ \bibinfo {pages} {125} (\bibinfo {year}
  {2008})}\BibitemShut {NoStop}%
\bibitem [{\citenamefont {Knysh}\ \emph
  {et~al.}(2011{\natexlab{a}})\citenamefont {Knysh}, \citenamefont
  {Smelyanskiy},\ and\ \citenamefont {Durkin}}]{Fupper}%
  \BibitemOpen
  \bibfield  {author} {\bibinfo {author} {\bibfnamefont {S.}~\bibnamefont
  {Knysh}}, \bibinfo {author} {\bibfnamefont {V.~N.}\ \bibnamefont
  {Smelyanskiy}}, \ and\ \bibinfo {author} {\bibfnamefont {G.~A.}\ \bibnamefont
  {Durkin}},\ }\href {\doibase 10.1103/PhysRevA.83.021804} {\bibfield
  {journal} {\bibinfo  {journal} {Phys. Rev. A}\ }\textbf {\bibinfo {volume}
  {83}},\ \bibinfo {pages} {021804} (\bibinfo {year}
  {2011}{\natexlab{a}})}\BibitemShut {NoStop}%
\bibitem [{\citenamefont {Lang}\ and\ \citenamefont
  {Caves}(2014)}]{lang2014optimal}%
  \BibitemOpen
  \bibfield  {author} {\bibinfo {author} {\bibfnamefont {M.~D.}\ \bibnamefont
  {Lang}}\ and\ \bibinfo {author} {\bibfnamefont {C.~M.}\ \bibnamefont
  {Caves}},\ }\href {\doibase 10.1103/PhysRevA.90.025802} {\bibfield  {journal}
  {\bibinfo  {journal} {Phys. Rev. A}\ }\textbf {\bibinfo {volume} {90}},\
  \bibinfo {pages} {025802} (\bibinfo {year} {2014})}\BibitemShut {NoStop}%
\bibitem [{\citenamefont {Sahota}\ and\ \citenamefont
  {Quesada}(2015)}]{sahota2015quantum}%
  \BibitemOpen
  \bibfield  {author} {\bibinfo {author} {\bibfnamefont {J.}~\bibnamefont
  {Sahota}}\ and\ \bibinfo {author} {\bibfnamefont {N.}~\bibnamefont
  {Quesada}},\ }\href {\doibase 10.1103/PhysRevA.91.013808} {\bibfield
  {journal} {\bibinfo  {journal} {Phys. Rev. A}\ }\textbf {\bibinfo {volume}
  {91}},\ \bibinfo {pages} {013808} (\bibinfo {year} {2015})}\BibitemShut
  {NoStop}%
\bibitem [{\citenamefont {{\v{S}}afr{\'a}nek}\ \emph
  {et~al.}(2015)\citenamefont {{\v{S}}afr{\'a}nek}, \citenamefont {Lee},\ and\
  \citenamefont {Fuentes}}]{vsafranek2015quantum}%
  \BibitemOpen
  \bibfield  {author} {\bibinfo {author} {\bibfnamefont {D.}~\bibnamefont
  {{\v{S}}afr{\'a}nek}}, \bibinfo {author} {\bibfnamefont {A.~R.}\ \bibnamefont
  {Lee}}, \ and\ \bibinfo {author} {\bibfnamefont {I.}~\bibnamefont
  {Fuentes}},\ }\href
  {http://iopscience.iop.org/article/10.1088/1367-2630/17/7/073016/meta}
  {\bibfield  {journal} {\bibinfo  {journal} {New Journal of Physics}\ }\textbf
  {\bibinfo {volume} {17}},\ \bibinfo {pages} {073016} (\bibinfo {year}
  {2015})}\BibitemShut {NoStop}%
\bibitem [{\citenamefont {Zhang}\ \emph {et~al.}(2013)\citenamefont {Zhang},
  \citenamefont {Yang},\ and\ \citenamefont {Wang}}]{zhang2013lossy}%
  \BibitemOpen
  \bibfield  {author} {\bibinfo {author} {\bibfnamefont {X.-X.}\ \bibnamefont
  {Zhang}}, \bibinfo {author} {\bibfnamefont {Y.-X.}\ \bibnamefont {Yang}}, \
  and\ \bibinfo {author} {\bibfnamefont {X.-B.}\ \bibnamefont {Wang}},\ }\href
  {\doibase 10.1103/PhysRevA.88.013838} {\bibfield  {journal} {\bibinfo
  {journal} {Phys. Rev. A}\ }\textbf {\bibinfo {volume} {88}},\ \bibinfo
  {pages} {013838} (\bibinfo {year} {2013})}\BibitemShut {NoStop}%
\bibitem [{\citenamefont {Adesso}(2014)}]{Adesso:2014hx}%
  \BibitemOpen
  \bibfield  {author} {\bibinfo {author} {\bibfnamefont {G.}~\bibnamefont
  {Adesso}},\ }\href {http://link.aps.org/doi/10.1103/PhysRevA.90.022321}
  {\bibfield  {journal} {\bibinfo  {journal} {Physical Review A}\ }\textbf
  {\bibinfo {volume} {90}},\ \bibinfo {pages} {022321} (\bibinfo {year}
  {2014})}\BibitemShut {NoStop}%
\bibitem [{\citenamefont {Anisimov}\ \emph {et~al.}(2010)\citenamefont
  {Anisimov}, \citenamefont {Raterman}, \citenamefont {Chiruvelli},
  \citenamefont {Plick}, \citenamefont {Huver}, \citenamefont {Lee},\ and\
  \citenamefont {Dowling}}]{Anisimov:2010kz}%
  \BibitemOpen
  \bibfield  {author} {\bibinfo {author} {\bibfnamefont {P.~M.}\ \bibnamefont
  {Anisimov}}, \bibinfo {author} {\bibfnamefont {G.~M.}\ \bibnamefont
  {Raterman}}, \bibinfo {author} {\bibfnamefont {A.}~\bibnamefont
  {Chiruvelli}}, \bibinfo {author} {\bibfnamefont {W.~N.}\ \bibnamefont
  {Plick}}, \bibinfo {author} {\bibfnamefont {S.~D.}\ \bibnamefont {Huver}},
  \bibinfo {author} {\bibfnamefont {H.}~\bibnamefont {Lee}}, \ and\ \bibinfo
  {author} {\bibfnamefont {J.~P.}\ \bibnamefont {Dowling}},\ }\href
  {http://link.aps.org/doi/10.1103/PhysRevLett.104.103602} {\bibfield
  {journal} {\bibinfo  {journal} {Phys. Rev. Lett.}\ }\textbf {\bibinfo
  {volume} {104}},\ \bibinfo {pages} {103602} (\bibinfo {year}
  {2010})}\BibitemShut {NoStop}%
\bibitem [{\citenamefont {Aspachs}\ \emph {et~al.}(2009)\citenamefont
  {Aspachs}, \citenamefont {Calsamiglia}, \citenamefont {Mu{\~n}oz-Tapia},\
  and\ \citenamefont {Bagan}}]{Aspachs:2009fu}%
  \BibitemOpen
  \bibfield  {author} {\bibinfo {author} {\bibfnamefont {M.}~\bibnamefont
  {Aspachs}}, \bibinfo {author} {\bibfnamefont {J.}~\bibnamefont
  {Calsamiglia}}, \bibinfo {author} {\bibfnamefont {R.}~\bibnamefont
  {Mu{\~n}oz-Tapia}}, \ and\ \bibinfo {author} {\bibfnamefont {E.}~\bibnamefont
  {Bagan}},\ }\href {http://link.aps.org/doi/10.1103/PhysRevA.79.033834}
  {\bibfield  {journal} {\bibinfo  {journal} {Physical Review A}\ }\textbf
  {\bibinfo {volume} {79}},\ \bibinfo {pages} {033834} (\bibinfo {year}
  {2009})}\BibitemShut {NoStop}%
\bibitem [{\citenamefont {Knysh}\ \emph
  {et~al.}(2011{\natexlab{b}})\citenamefont {Knysh}, \citenamefont
  {Smelyanskiy},\ and\ \citenamefont {Durkin}}]{Knysh:2011ed}%
  \BibitemOpen
  \bibfield  {author} {\bibinfo {author} {\bibfnamefont {S.}~\bibnamefont
  {Knysh}}, \bibinfo {author} {\bibfnamefont {V.~N.}\ \bibnamefont
  {Smelyanskiy}}, \ and\ \bibinfo {author} {\bibfnamefont {G.~A.}\ \bibnamefont
  {Durkin}},\ }\href {http://link.aps.org/doi/10.1103/PhysRevA.83.021804}
  {\bibfield  {journal} {\bibinfo  {journal} {Physical Review A}\ }\textbf
  {\bibinfo {volume} {83}},\ \bibinfo {pages} {021804} (\bibinfo {year}
  {2011}{\natexlab{b}})}\BibitemShut {NoStop}%
\bibitem [{\citenamefont {Ko{\l}ody{\'{n}}ski}\ and\ \citenamefont
  {Demkowicz-Dobrza{\'{n}}ski}(2010)}]{Koiodynski:2010cg}%
  \BibitemOpen
  \bibfield  {author} {\bibinfo {author} {\bibfnamefont {J.}~\bibnamefont
  {Ko{\l}ody{\'{n}}ski}}\ and\ \bibinfo {author} {\bibfnamefont
  {R.}~\bibnamefont {Demkowicz-Dobrza{\'{n}}ski}},\ }\href
  {http://link.aps.org/doi/10.1103/PhysRevA.82.053804} {\bibfield  {journal}
  {\bibinfo  {journal} {Physical Review A}\ }\textbf {\bibinfo {volume} {82}},\
  \bibinfo {pages} {053804} (\bibinfo {year} {2010})}\BibitemShut {NoStop}%
\bibitem [{\citenamefont {Escher}\ \emph {et~al.}(2011)\citenamefont {Escher},
  \citenamefont {de~Matos~Filho},\ and\ \citenamefont
  {Davidovich}}]{Escher:2011fn}%
  \BibitemOpen
  \bibfield  {author} {\bibinfo {author} {\bibfnamefont {B.~M.}\ \bibnamefont
  {Escher}}, \bibinfo {author} {\bibfnamefont {R.~L.}\ \bibnamefont
  {de~Matos~Filho}}, \ and\ \bibinfo {author} {\bibfnamefont {L.}~\bibnamefont
  {Davidovich}},\ }\href@noop {} {\bibfield  {journal} {\bibinfo  {journal}
  {Nature Physics}\ }\textbf {\bibinfo {volume} {7}},\ \bibinfo {pages} {406}
  (\bibinfo {year} {2011})}\BibitemShut {NoStop}%
\bibitem [{\citenamefont {Demkowicz-Dobrza{\'{n}}ski}\ \emph
  {et~al.}(2012)\citenamefont {Demkowicz-Dobrza{\'{n}}ski}, \citenamefont
  {Ko{\l}ody{\'{n}}ski},\ and\ \citenamefont
  {Gu{\c{t}}{\u{a}}}}]{DemkowiczDobrzanski:2012gl}%
  \BibitemOpen
  \bibfield  {author} {\bibinfo {author} {\bibfnamefont {R.}~\bibnamefont
  {Demkowicz-Dobrza{\'{n}}ski}}, \bibinfo {author} {\bibfnamefont
  {J.}~\bibnamefont {Ko{\l}ody{\'{n}}ski}}, \ and\ \bibinfo {author}
  {\bibfnamefont {M.}~\bibnamefont {Gu{\c{t}}{\u{a}}}},\ }\href@noop {}
  {\bibfield  {journal} {\bibinfo  {journal} {Nature Communications}\ }\textbf
  {\bibinfo {volume} {3}},\ \bibinfo {pages} {1063} (\bibinfo {year}
  {2012})}\BibitemShut {NoStop}%
\bibitem [{\citenamefont {Jarzyna}\ and\ \citenamefont
  {Zwierz}(2015)}]{PhysRevA.92.032112}%
  \BibitemOpen
  \bibfield  {author} {\bibinfo {author} {\bibfnamefont {M.}~\bibnamefont
  {Jarzyna}}\ and\ \bibinfo {author} {\bibfnamefont {M.}~\bibnamefont
  {Zwierz}},\ }\href {\doibase 10.1103/PhysRevA.92.032112} {\bibfield
  {journal} {\bibinfo  {journal} {Phys. Rev. A}\ }\textbf {\bibinfo {volume}
  {92}},\ \bibinfo {pages} {032112} (\bibinfo {year} {2015})}\BibitemShut
  {NoStop}%
\bibitem [{\citenamefont {Braunstein}\ and\ \citenamefont
  {Caves}(1994)}]{braunstein94a}%
  \BibitemOpen
  \bibfield  {author} {\bibinfo {author} {\bibfnamefont {S.~L.}\ \bibnamefont
  {Braunstein}}\ and\ \bibinfo {author} {\bibfnamefont {C.~M.}\ \bibnamefont
  {Caves}},\ }\href {\doibase 10.1103/PhysRevLett.72.3439} {\bibfield
  {journal} {\bibinfo  {journal} {Phys. Rev. Lett.}\ }\textbf {\bibinfo
  {volume} {72}},\ \bibinfo {pages} {3439} (\bibinfo {year}
  {1994})}\BibitemShut {NoStop}%
\bibitem [{\citenamefont {Braunstein}\ \emph {et~al.}(1996)\citenamefont
  {Braunstein}, \citenamefont {Caves},\ and\ \citenamefont
  {Milburn}}]{braunstein96a}%
  \BibitemOpen
  \bibfield  {author} {\bibinfo {author} {\bibfnamefont {S.~L.}\ \bibnamefont
  {Braunstein}}, \bibinfo {author} {\bibfnamefont {C.~M.}\ \bibnamefont
  {Caves}}, \ and\ \bibinfo {author} {\bibfnamefont {G.~J.}\ \bibnamefont
  {Milburn}},\ }\href {\doibase 10.1006/aphy.1996.0040} {\bibfield  {journal}
  {\bibinfo  {journal} {Ann. Phys.}\ }\textbf {\bibinfo {volume} {247}},\
  \bibinfo {pages} {135} (\bibinfo {year} {1996})}\BibitemShut {NoStop}%
\bibitem [{\citenamefont {Helstrom}(1976)}]{helstrom76a}%
  \BibitemOpen
  \bibfield  {author} {\bibinfo {author} {\bibfnamefont {C.~W.}\ \bibnamefont
  {Helstrom}},\ }\href@noop {} {\emph {\bibinfo {title} {Quantum detection and
  estimation theory}}},\ \bibinfo {edition} {1st}\ ed.,\ \bibinfo {series}
  {Mathematics in science and engineering}, Vol.\ \bibinfo {volume} {123}\
  (\bibinfo  {publisher} {Academic Press},\ \bibinfo {address} {New York},\
  \bibinfo {year} {1976})\BibitemShut {NoStop}%
\bibitem [{\citenamefont {Holevo}(1982)}]{holevo82a}%
  \BibitemOpen
  \bibfield  {author} {\bibinfo {author} {\bibfnamefont {A.~S.}\ \bibnamefont
  {Holevo}},\ }\href@noop {} {\emph {\bibinfo {title} {Probabilistic and
  statistical aspects of quantum theory}}},\ \bibinfo {edition} {1st}\ ed.,\
  \bibinfo {series} {North-Holland series in statistics and Probability
  theory}, Vol.~\bibinfo {volume} {1}\ (\bibinfo  {publisher} {North-Holland},\
  \bibinfo {address} {Amsterdam},\ \bibinfo {year} {1982})\BibitemShut
  {NoStop}%
\bibitem [{\citenamefont {Hayashi}(2006)}]{hayashi2006quantum}%
  \BibitemOpen
  \bibfield  {author} {\bibinfo {author} {\bibfnamefont {M.}~\bibnamefont
  {Hayashi}},\ }\href@noop {} {\emph {\bibinfo {title} {Quantum Information}}}\
  (\bibinfo  {publisher} {Springer},\ \bibinfo {year} {2006})\BibitemShut
  {NoStop}%
\bibitem [{\citenamefont {Marian}\ and\ \citenamefont
  {Marian}(2012)}]{marian2012uhlmann}%
  \BibitemOpen
  \bibfield  {author} {\bibinfo {author} {\bibfnamefont {P.}~\bibnamefont
  {Marian}}\ and\ \bibinfo {author} {\bibfnamefont {T.~A.}\ \bibnamefont
  {Marian}},\ }\href {\doibase 10.1103/PhysRevA.86.022340} {\bibfield
  {journal} {\bibinfo  {journal} {Phys. Rev. A}\ }\textbf {\bibinfo {volume}
  {86}},\ \bibinfo {pages} {022340} (\bibinfo {year} {2012})}\BibitemShut
  {NoStop}%
\bibitem [{\citenamefont {Demkowicz-Dobrza{\'{n}}ski}\ \emph
  {et~al.}(2013)\citenamefont {Demkowicz-Dobrza{\'{n}}ski}, \citenamefont
  {Banaszek},\ and\ \citenamefont {Schnabel}}]{DemkowiczDobrzanski:2013fs}%
  \BibitemOpen
  \bibfield  {author} {\bibinfo {author} {\bibfnamefont {R.}~\bibnamefont
  {Demkowicz-Dobrza{\'{n}}ski}}, \bibinfo {author} {\bibfnamefont
  {K.}~\bibnamefont {Banaszek}}, \ and\ \bibinfo {author} {\bibfnamefont
  {R.}~\bibnamefont {Schnabel}},\ }\href
  {http://link.aps.org/doi/10.1103/PhysRevA.88.041802} {\bibfield  {journal}
  {\bibinfo  {journal} {Physical Review A}\ }\textbf {\bibinfo {volume} {88}},\
  \bibinfo {pages} {041802} (\bibinfo {year} {2013})}\BibitemShut {NoStop}%
\bibitem [{\citenamefont {Shaji}\ and\ \citenamefont
  {Caves}(2007)}]{Shaji:2007bt}%
  \BibitemOpen
  \bibfield  {author} {\bibinfo {author} {\bibfnamefont {A.}~\bibnamefont
  {Shaji}}\ and\ \bibinfo {author} {\bibfnamefont {C.~M.}\ \bibnamefont
  {Caves}},\ }\href {http://link.aps.org/doi/10.1103/PhysRevA.76.032111}
  {\bibfield  {journal} {\bibinfo  {journal} {Physical Review A}\ }\textbf
  {\bibinfo {volume} {76}},\ \bibinfo {pages} {032111} (\bibinfo {year}
  {2007})}\BibitemShut {NoStop}%
\end{thebibliography}%
\end{document}